\documentclass[useAMS,usenatbib]{mn2e} \input{epsf}

\def\gsim{ \lower .75ex \hbox{$\sim$} \llap{\raise .27ex \hbox{$>$}} }
\def\lsim{ \lower .75ex \hbox{$\sim$} \llap{\raise .27ex \hbox{$<$}} }

\title[The interacting LMC-MW system] {The gravitational and hydrodynamical interaction between the LMC and the Galaxy}

\author[C. Mastropietro et al.]  {C. Mastropietro,$^{1}$\thanks{E-mail:
chiara@physik.unizh.ch} B. Moore$^{1}$, L. Mayer$^{1}$, J. Wadsley$^{2}$ and
J. Stadel$^{1}$\\ $^1$Institute for Theoretical Physics, University of Z\"urich,
CH-8057 Z\"urich, Switzerland\\ $^2$Department of Physics \& Astronomy, McMaster
University, 1280 Main St. West, Hamilton ON L8S 4M1 Canada} 

\begin{document}


\pagerange{\pageref{firstpage}--\pageref{lastpage}} \pubyear{00}

\maketitle

\label{firstpage}

\begin{abstract} We use high resolution N-Body/SPH simulations to study the
hydrodynamical and gravitational interaction between the Large Magellanic Cloud
and the Milky Way.  We model the dark and hot extended halo components as well
as the stellar/gaseous disks of the two galaxies. Both galaxies are embedded in
extended cuspy LCDM dark matter halos. We follow the previous four Gyrs of
the LMC's orbit such that it ends up with the correct location and orientation
on the sky. Tidal forces elongate the LMC's disk, forcing 
a bar and creating a strong warp and diffuse stellar halo, although very few
stars become unbound.  The stellar halo may account for some of the 
microlensing events.
Ram-pressure from a low density ionised halo is then
sufficient to remove $1.4 \times 10^8M_\odot$ of gas from the LMC's disk forming
a great circle trailing stream around the Galaxy. The column density of stripped
gas falls by two orders of magnitude 100 degrees from LMC. The LMC does not induce any
response in the Milky Way disk. On the contrary, the tides raised by the Milky Way 
determine the truncation of the satellite at about 11 kpc. After several Gyrs of interaction
the gas disk of the LMC is smaller than the stellar disk due to ram pressure 
and its size compares well with the observational data.

\end{abstract}


\section{Introduction}

The Large Magellanic Cloud (LMC) is in the process of destroying its neighbour,
the Small Magellanic Cloud (SMC).  In turn, the Milky Way (MW) is playing a role
reshaping these galaxies and creating the spectacular Magellanic Stream (hereafter MS).  
The MS is a trailing filament of neutral hydrogen
that originates from the Magellanic Clouds and stretches for over $\sim 100$ degrees in 
the Southern Sky and is the most prominent signature of an interaction between the
Clouds and the Milky Way. The combined interaction between these three galaxies
must strongly influence the morphological evolution of the Clouds, as well as
their internal kinematics and star formation history. A resonant interaction
between the Galaxy and the LMC could even be sufficient to excite the warp
observed in the Galactic disk \citep{Weinberg95, Weinberg98}. \\

Several models have been proposed in order to study the nature of
this interaction and to isolate the main mechanism responsible for the formation
of the Stream. These models have become increasingly refined over the years.
The first gravitational tidal models took into account just the binary
interaction between the LMC and MW \citep{Lin77, Murai}, but later models
included a close interaction between the Clouds themselves during a previous
passage at the perigalacticon about 1.5 Gyrs ago \citep{Gardiner94, Lin95,
Gardiner96}.

The strongest argument which supports a tidal origin for the MS is provided by
the discovery of neutral hydrogen in front of the LMC \citep{PutmanNature}. However the amount
of material in this feature is just a fraction of the mass of the MS
\citep{Putman03}. However there are some characteristics of the Stream that
tidal models cannot explain. In particular they fail in reproducing the gradual
decrease of the HI column density and the lack of a
stellar tidal feature \citep{Mathewson79, Brueck, Guhathakurta}. Stars are
expected to be stripped by gravitational forces along with the gas, although
\citet{Weinberg00} suggests that the stellar debris would have a different
distribution respect to the gaseous Stream and form a diffuse envelope around the
LMC. Traces of an offset old stellar stream could be identified with the giant
stars observed by \citet{Majewski} around the Magellanic Clouds.

A hydrodynamical interaction has also been invoked to explain the Stream.
\citet{Mathewson87} proposed that gas is swept away from the intra-cloud region
and decelerated by random collisions with high velocity clouds (HVC). This model
requires unrealistic densities for the HVCs in the MW halo in order to justify
the high number of encounters that should generate the Stream and does not
explain the column density gradient. Continuous stripping 
models \citep{Meurer, Sofue, Moore94} propose a scenario 
in which gas is stripped by ram-pressure forces from
the outer regions of the Clouds through an interaction with a distribution of
hot gas in the halo or an extended ionised disk. A satellite moving with a
relative velocity $v$ respect to this external medium of density $\rho$
experiences a pressure $P=\rho {v}^2$ which can be sufficient to
remove part of the neutral gas from the disk.

Together with the difficulties in creating a continuous leading arm -- although
the stripped material decelerated by ram pressure forces could fall on smaller
orbits and lead the Clouds \citep{Sofue, Moore94} -- ram pressure models are
thought to be able to reproduce the large amount of gas observed in the Stream
($\sim 2\times 10^8 M_{\odot}$ according to Putman et al. 2003a).\\ The presence
of an extended hot halo surrounding the MW and in hydrostatic equilibrium within
the dark matter potential is expected by current models of hierarchical
structure formation \citep{White91}. This gaseous halo, left over from the
initial collapse, would be continuously fuelled by accretion of satellite
galaxies over time. Recent observational results have confirmed the existence of
a distribution of hot ($T$ $\sim 10^6$ K) gas well beyond the Galactic disk, with
a radius larger than 70 kpc in order to explain some ionisation features
discovered in the MS \citep{Sembach, Putman03halo}. Constraints from dynamical
and thermal arguments fix the density of the gaseous halo in a range between
$10^{-5}$ and $10^{-4} \textrm{cm}^{-3}$ at the LMC distance from the Galactic
center. The Clouds are therefore subjected along their orbits to the ram
pressure generated by a tenuous distribution of hot gas and the Magellanic
System itself seems to be the result of a complex interaction that involves both
hydrodynamical and tidal processes.\\

The aim of this work is to study for the first time the simultaneous effect of
gravitational and hydrodynamical forces acting on the LMC as it moves within the
Galactic halo, using fully self consistent galaxy models. In particular we are
interested in the formation and evolution of the Magellanic 
Stream and in the dynamical
changes in the internal structure of the LMC due to the interaction with the
MW. We ignore the SMC owing to its small mass ($\sim 10 \%$ of the LMC),
neglecting the possibility of a close encounter between the two Clouds and the
consequence of additional loss of mass. This approximation would lead to an
underestimation of the amount of HI in the Stream.  \\

This paper is structured as follows. In Section 2 we present the main
characteristics of the galaxy models that we use. The results of test wind tube
simulations are discussed in Section 3. In Section 4 we analyse the results of the interacting runs regarding the Stream and the LMC disk. 
  
\section{Galaxy Models}

The initial conditions of the simulations are constructed using the technique
described by \citet{Hernquist}. Both the Milky Way and the Large Magellanic
Cloud are multi-component systems with a stellar and gaseous disk embedded in a
spherical dark matter halo. The density profile of the NFW \citep{Navarro97} 
halo is adiabatically contracted due to baryonic cooling
\citep{Springel}. Stars and cold ($T=10^4$K) gas in the disk follow the same
exponential surface density profile of the form 
\begin{equation} \label{disk}
\Sigma(R)=\frac{M_d}{2\pi {R_d}^2}\,\textrm{exp}\,(-R/R_d)\, , \end{equation}
where $M_d$ and $R_d$ are the disk mass and radial scale length
(in cylindrical coordinates) respectively, while the thin vertical structure is characterised by the scale height $z_d \sim \frac{1}{5}R_d$: 
\begin{equation} \label{diskh}
\rho_d(R,z) = \frac{\Sigma(R)}{2z_d}\,\textrm{sech}^2\,(z/z_d)\, .
\end{equation} 
The MW model comprises of two further components: a stellar bulge
and an extended hot gaseous halo. 
The bulge has a spherical Hernquist
\citep{Hernquistan} profile: \begin{equation} \label{bulge} \rho_b(r) =
\frac{M_b}{2\pi} \frac{r_b}{r(r_b+r)^3} 
\end{equation}
where $M_b$ is the bulge
mass and $r_b$ its scale length.

\begin{table} \caption{Galaxy models: for each component mass (in units of
$10^{10} M_{\odot}$ and scale radius (kpc) are indicated. The last three columns
refer to high resolution runs and report number of particles, mass particle (in
units of $10^5 M_{\odot}$) and softening length (kpc).}
\begin{tabular}{l|c|c|c|c|c|} \hline
 MW &Mass &Scale radius & $N$ & $m$ & $\epsilon$\\ \hline Bulge & 0.73 & 0.70 &
 $10^4$ & 7.3 & 0.1 \\ Gaseous disk & 0.46 & 3.53 & $10^5$ & 0.46 & 0.1 \\
 Stellar disk & 4.10 & 3.53 & $10^5$ & 4.10 & 0.1\\ DM halo & 104.0 & 18.2 &
 $10^6$ & 10.4 & 0.5\\ Hot halo & 1.0 & 18.2 & $5 \times 10^5$ & 0.20 & 0.5\\
 \hline LMC &Mass &Scale radius & N & m & $\epsilon$\\ \hline Gaseous disk & 0.11
 & 1.66 & $10^5$ & 0.11 & 0.1 \\ Stellar disk & 0.11 & 1.66 & $5 \times 10^4$ &
 0.22 & 0.1\\ DM halo & 2.38 & 6.02 & $6 \times 10^5$ & 10.4 & 0.5\\

\hline

\end{tabular} \end{table}

Although the presence of an extended corona of hot gas seems to be the most
likely explanation for the detection of OVI and OVII absorption lines associated 
with HI structures (MS, some HVCs, Outer Spiral Arm, Complex A, Complex C, 
according to Sembach et al. 2003), its density structure is still uncertain,
particularly at large distances from the Galactic center. These ionization features
suggest a mean density of the gaseus halo within 80 kpc of 
$\sim 10^{-4} - 10^{-5} \textrm{cm}^{-3}$ \citep{Sembach03}. \citet{Stanimirovic} provides similar values for the density
at 45 kpc ($n_h \, \lsim 3 \times 10^{-4} \textrm{cm}^{-3}$) in order to
explain the confinement mechanism of the MS clouds through pressure support from
the hot halo. Another constraint comes from the detection of H$\alpha$ emission
along the MS which cannot be explained solely by photoionisation from the
Galactic disk \citep{Putman03halo} and could be induced by the interaction with
a relative dense ($n_h > 10^{-4} \textrm{cm}^{-3}$ at 50 kpc) gaseous
environment \citep{Weiner}. With a density higher than $10^{-4} \textrm{cm}^{-3}$
\citet{Quilis01} are able to reproduce, using hydrodynamical simulations, the
head-tail structure and the compression fronts associated with HVCs, in the case
of either pure gas clouds or gas embedded within a dark matter halo. A
significantly lower value for the upper limit of the halo density ($n_h
<10^{-5} \textrm{cm}^{-3}$) is found by \citet{Murali} by requiring the survival
of the cloud MS IV for 500 Myrs against evaporation. However we
emphasise the fact that the evaporation time strongly depends on the temperature of the
gaseous halo \citep{Cowie} and a choice of $T\sim10^6 \textrm{K}$, three times
lower than the value used by Murali, increases by almost a factor fifteen the
lifetime of MS IV, making the survival of the cloud possible at a
density of even $10^{-4} \textrm{cm}^{-3}$. At larger distances from the Galactic
center constraints on the halo density are poor, although
\citet{Blitz} derive a value of $2.5 \times 10^{-5} \textrm{cm}^{-3}$ at a
distance of 200 kpc by applying ram pressure stripping to dwarf spheroidal
galaxies within the Galactic halo.

We model the hot halo with a spherical distribution of gas which traces the dark
matter profile within the virial radius and is in hydrostatical equilibrium
inside the Galactic potential. Assuming an isotropic model, the halo temperature
at a given radius $r$ is determined by the cumulative mass distribution $M(r)$ of
the dark, stellar and gaseous components of the MW beyond $r$ and by the density
profile $\rho_h(r)$ of the hot gas: 
\begin{equation}\label{gashalo} 
T(r) = \frac{m_p}{k_B} \frac{1}{\rho_h(r)} \int_{r}^{\infty} \rho_h(r)\frac{GM(r)}{r^2} \, dr \, , 
\end{equation} where $m_p$ is the proton mass, $G$
and $k_B$ are the gravitational and Boltzmann constants. The halo gas is
completely ionised at a mean temperature of $10^6$ K. We normalise the density
profile using a hot halo mass of $M_h = 10^{10} M_{\odot}$, which
corresponds to a mean density of $2 \times 10^{-5} \textrm{cm}^{-3}$ within 150
kpc and $ 8.5 \times 10^{-5}$ at 50 kpc. With the assumption that the
gas traces the dark matter we overestimate by a factor of
three the ionised gas density near the disk, however this is not important
for the current study.
\begin{figure}
\epsfxsize=8truecm \epsfbox{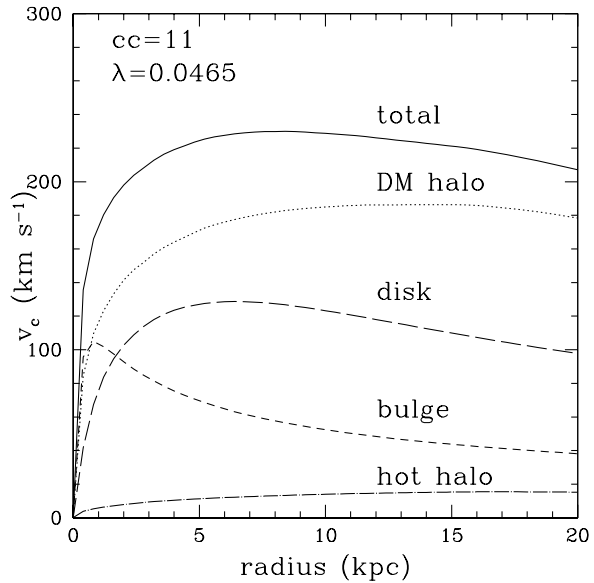} \caption{MW rotation curves. On the top left spin and concentration parameter are indicated.}
\end{figure}

\begin{figure} 
\epsfxsize=8truecm \epsfbox{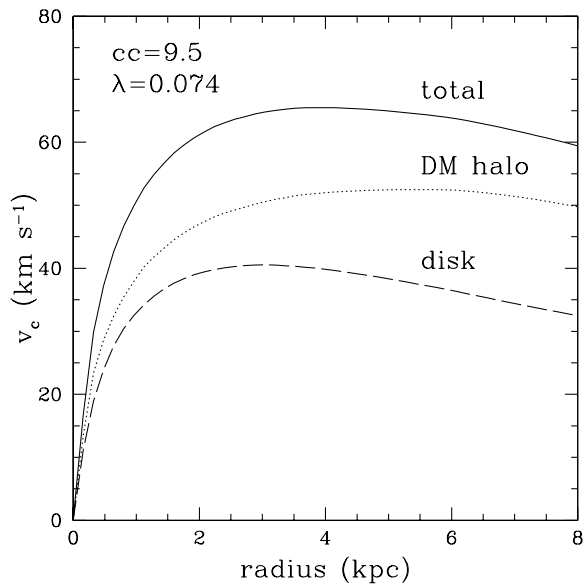} \caption{LMC
rotation curves. On the top left spin and concentration parameter are
indicated. } 
\end{figure} 
The masses and the scale lengths of the MW and LMC
models (Table 1) are selected in order to reproduce observational
constraints from \citet{Kim} and \citet{vanderMarel02}, while for the MW we
adopt a model similar to model $A_1$ of \citet{Klypin}.  Fig. 1 and Fig. 2 
show the rotation
curves of the different components of the two galaxies. The concentration $c$ and
the spin parameter $\lambda $ \citep{Mo} are indicated at the top left. The
concentration is defined as $ cc = r_{vir}/r_s$, where $r_{vir} $ and $r_s$ are
the virial and the scale radius of the NFW halo, whereas the spin parameter
relates the angular momentum J and the total energy E of an halo with a virial
mass $M_{vir}$ through the relation $\lambda = J|E|^{1/2}G^{-1}M_{vir}^{-5/2}$.
The disk of the satellite galaxy has, within the scale radius, a central mass
surface density of $\sim 70\, \textrm{M}_{\odot} \textrm{pc}^{-2}$, which
corresponds to a B-band surface brightness of $\sim 23\, \textrm{mag
arcsec}^{-2}$, if we adopt a mass to light ratio of two.

The thickness of the disks is set such that Toomre's
\citep{Toomre64} stability criterion is satisfied, which requires, for a stellar disk
\begin{equation} \label{Toomres} Q_{star}(r)= \frac{\sigma_R k}{3.36 G
\Sigma_s}> 1,
\end{equation} where $\sigma_R$ is the radial velocity dispersion,
$k$ is the local epicyclic frequency and $\Sigma_s$ the stellar surface
density. For a gaseous disk Equation \ref{Toomres} becomes: 
\begin{equation}
\label{Toomreg} Q_{gas}(r)= \frac{v_s k}{\pi\, G \Sigma_g} > 1,
\end{equation} where $v_s$ is the sound velocity and $\Sigma_g$ the surface
density of the gas.

All the simulations we now discuss were 
carried out using GASOLINE, a parallel tree-code with multi-stepping
\citep{Wadsley}.  The high resolution runs have $2.46 \times 10^6$ particles, of
which $3.5 \times 10^5$ are used for the disks and $5 \times 10^5$ for the hot
halo of the MW (Table 1). The gravitational spline softening is set equal to 0.5 kpc
for the dark and gaseous halos, and to 0.1 kpc for stars and gas in the disk and
bulge components. \\

\section{Test simulations}

The mass resolution of the simulations strongly influences the ram pressure
process. In particular, if the particles of the hot external medium are too
massive, they do not flow smoothly on to the disk, but produce scattering and numerical
holes, increasing the stripping process which is dominated by noise. In order
to study these effects, we perform several test simulations in which
we place the model LMC galaxy within a ``wind tunnel'' and 
vary the mass of the hot halo particles for a fixed resolution 
of the LMC disk.

The first set (H1, H2 and H3) of runs simulates the passage of the satellite at
the perigalacticon, which is the point along the orbit where ram
pressure is most efficient due to the high values of $\rho_g$ and $v$. The volume
of the simulations is an oblong of base equal to the diameter of the LMC and
height $h=vt$, where $v$ is the velocity of the satellite at the perigalacticon
($250 \,\textrm{km}\,\textrm{s}^{-1}$) and $t$ is set $\sim 2$ Gyrs ($\sim$ the
inferred orbital time for the LMC). We represent the hot gas as a flux of
particles moving with a velocity $v$ along the $z$ axis, parallel to the angular
momentum of the disk. The hot particles have an initial random distribution, a
temperature $T=10^6 \textrm{K}$ and a number density $n_h=8\times 10^{-5}
\textrm{cm}^{-3}$. The box has periodic boundary conditions in order to restore
the flow of hot gas that leaves the oblong. The disk particles have the same
mass $m_{disk}$ as in the high resolution interacting runs (see Table 2),
while the mass ratio with the particles in the oblong
$m_h/m_{disk}$ of 1:10, 1:2 and 1:1.
 Gas is removed from the disk if the ram pressure produced by the hot halo of
 the MW is greater than the restoring force per unit area provided by the disk
 of the satellite. The condition for ram pressure
 stripping is expressed by \citep{Gunn} 
\begin{equation} 
\rho_h v^2 > 2 \pi G \Sigma(R) \Sigma_g(R), 
\end{equation} where $v$ is the velocity of the galaxy
 with respect to the surrounding medium, $\rho_h$ is the density of the hot halo
 of the MW and $\Sigma_g(r)$ is the cold gas surface density at the radius
 $R$. $\Sigma$ represents the gravitational surface mass density of the
 disk. The minimum radius given by Equation (7) is the final stripping radius
 $R_{str}$ beyond which the ISM can be removed from the galaxy.
 The halo component does not contribute to the
 restoring force initially, since in the case of a face on impact any force
 associated to a spherical potential is perpendicular to the wind pressure. As
 the gas is stripped out from the disk in the $z$ direction, the disk gravity
 decreases while the projection of the radial halo gravity force in the $z$
 direction rises \citep {Schulz}. This effect is particularly important in the
 case of low ram pressure regimes as the ones we are interested in here. The condition for stripping expressed in Equation (7) implicitly
 assumes instantaneous stripping from the disk and discrete ram pressure stripping
 events such that the loss of mass approaches an asymptotic limit. If instead ram
 pressure is a continuous
 process with time scales comparable with the orbital time of the satellite, the
 stripping radius calculated analytically underestimates the physical size of
 the gaseous disk. Such a continuous ram pressure stripping occurs here not 
because of turbulent or viscous stripping but because the gas at the edge
of the disk heats up responding to the compression exerted by the outer
medium and can therefore become unbound by increasing its energy despite 
the ram pressure wind being constant. Timescales of both Kelvin-Helmoltz
instabilities and artificial viscous forces are close to 10 Gyr because of,
respectively, stabilization due to the cuspy halo potential of the LMC
and the high mass resolution employed, and are therefore negligible over
the timescales explored here. 

We estimate $R_{str}$ as the radius which contains 99\% of the
 gas particles in the midplane of the disk. The mass of gas that remains in the
 disk can be computed summing the disk particles inside a cylinder of radius
 $R_{str}$ and thickness 1 kpc.   
The mass of gas bound to the satellite at a time $t$ is
 instead not directly related to $R_{str}$, but is determined by the
 potential of the dark halo. We calculate it balancing kinetic, thermal and
 potential energy in concentric spherical shells centered on the LMC, defining
 the gas tidal radius $r_{g}$ as the radius of the most distant bound shell.

 \begin{table*} \centering \begin{minipage}{140mm} 
\caption{LMC wind tunnel simulations. For each run the density $\rho_h$ of the hot medium and the
velocity $v$ of the satellite are indicated. The resolution of each model is
expressed in terms of the mass ratio $m_{h}/m_{disk}$ between
halo and disk particles. $N$ is the number of hot particles, while the last
four columns give the ram pressure stripping radius
$R_{str}$ and the mass stripped from the disk $m_{str}$, the
gas tidal radius $r_g$ and the mass of gas unbound to the galaxy
$m_{unb}$. In the case of edge on galaxies the stripping radius
$R_{str}$ is the average value between the two sides of the disk. All the values are calculated at $t=2$ Gyrs.}
\begin{tabular}{l|c|c|c|c|c|c|c|c|c} \hline
 Run & Model & $\rho_{g}$ & $v$& $m_{h}/m_{disk}$ & $N$
 & $R_{str}$ & $m_{str}$ & $r_{g}$ &
 $m_{unb}$ \\ & & ($10^{-5}\textrm{cm}^{-3}$)&
 ($\textrm{km}\,\textrm{s}^{-1}$) & & $(10^6)$ & (kpc) & ($10^8 M_\odot$) &
 (kpc) & ($10^8 M_\odot$)\\ \hline
 H1 &face on& 8& 250&10 & 0.37 & 4.0 & 4.11 &6.7 & 2.84\\
 H2 &face on& 8& 250&2 & 1.86 & 5.3 & 2.76 &7.5& 1.32\\
 H3 &face on& 8& 250&1 & 3.2 & 5.5 &2.52 &7.8& 1.0\\ \hline
 L1&face on& 3.5& 175&2 & 0.54 & 6.6 &2.09 &8.3&0.58\\
 L2&face on& 3.5& 175&1 & 1.07 & 6.8 & 2.02 &8.5& 0.44\\
 L3&face on& 3.5& 175&0.5 & 2.14 & 6.9 & 1.99 &8.6& 0.42\\ \hline
 E1&edge on & 8& 250& 2& 1.86 & 5.5 & 2.74 &7.7& 1.32\\
 E2&edge on& 3.5& 175& 2& 1.07 & 6.2 & 2.32 &8.1& 0.79\\ \hline \end{tabular}
 \end{minipage} \end{table*}

\begin{figure} \epsfxsize=8truecm \epsfbox{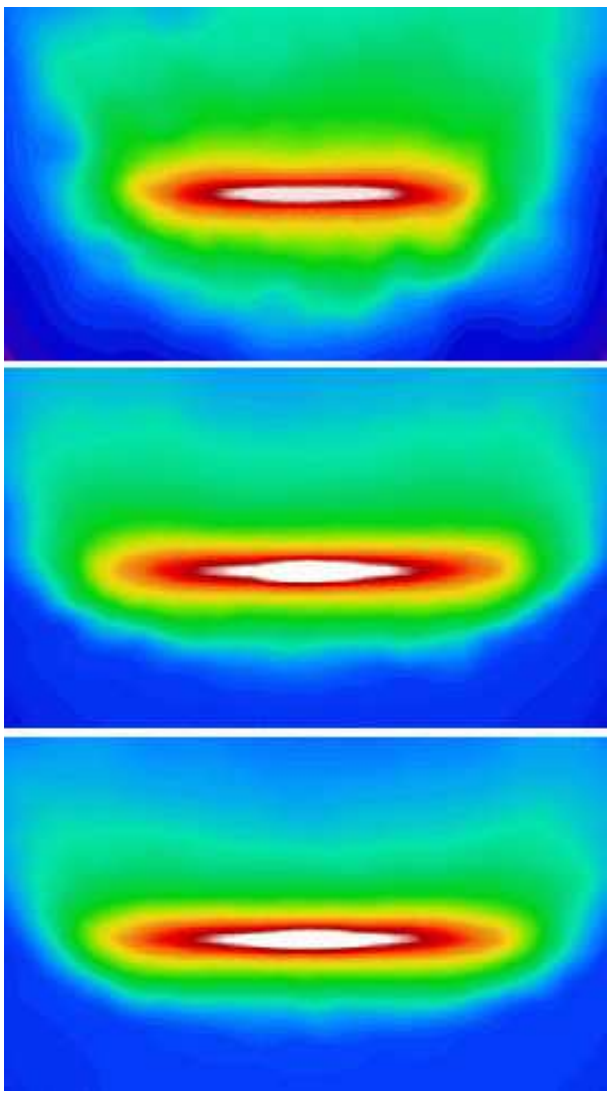} \caption{From the top to
the bottom: column density distribution for the LMC test models H1, H2, H3. The
scale is logarithmic with blue corresponding to the density of the hot halo
gas.}  \end{figure}

\begin{figure} \epsfxsize=8truecm \epsfbox{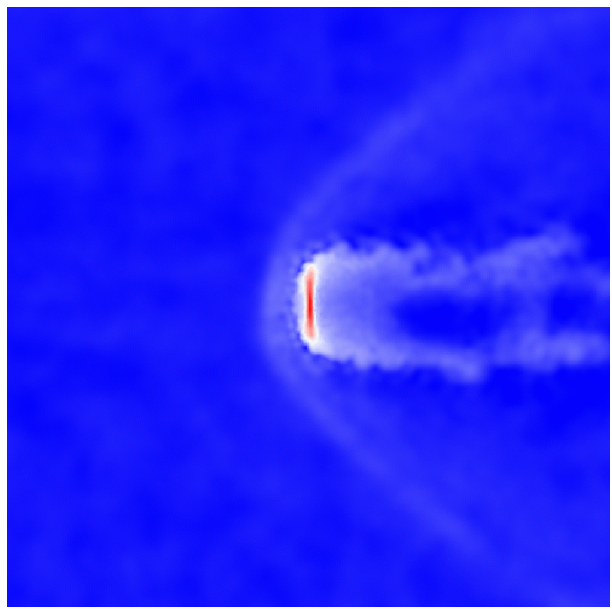} 
\caption{H3 run: density distribution of gas in a thin slice perpendicular to the LMC disk. 
The satellite is moving face on toward the left side of the picture while a shock front forms in 
the external medium in front of the disk. }  \end{figure}
 
The mass of cold gas stripped from the LMC's disk $m_{str}$ decreases as we
decrease the mass of the halo particles: in H1 $m_{str}$ is
$\sim 1.6 $ times larger than in H3 (Table 2). Mass resolution also determinates
the shape of the front edge and influences the morphology of the gaseous
disk. Fig. 3 shows the gas density in the disk after 2 Gyrs for the H1, H2 and H3
runs. Numerical effects in the low resolution run are quite evident: the massive
hot particles produce particle -- particle interactions with gas in the galaxy,
creating holes and discontinuities along the front edge of the disk. The spiral
pattern is partially destroyed and the disk looses its symmetry. Scattering
reduces the stripping radius by a factor of 1.4 and produces a sharp truncation,
while the effect of pure ram pressure is much more gradual, since halo particles
stream around the disk. Finally, the removed gas forms an asymmetric and
turbulent tail. The high resolution H3 run appears very symmetric 
with a flat and uniform edge and a regular stream, which flows around
the disk and creates a low density region behind (Fig. 4). As
previously noticed by several authors \citep {Quilis00, Quilis01, Schulz}, a bow
shock forms in front of the disk, increasing the local ICM temperature from
$10^6$ to $ 2.5 \times 10 ^6$ degrees. The intermediate run H2 is 
similar to H1, although the gas tail is more pronounced and the amount of
stripped gas is less than $ 10\%$ larger. The resolution (mass ratio of 1:2)
of the H2 run is 
to be a good compromise between the necessity of solving the gas dynamics
and the number of particles to use in a full production simulation. 

The runs L1, L2 and L3 have a gas density
($\rho_h =3.5 \times 10^{-5} \textrm{cm}^{-3}$) and satellite velocity (175
$\textrm{km}\,{\textrm{s}}^{-1}$) similar to our final LMC's orbit with
apocentric distance $\sim 50$ kpc and apo/peri ratio $\sim 2.5:1$ 
(see section 4.1). Reducing the ram pressure by
a factor of three decreases the stripped mass by $\sim
20\%$, while the amount of unbound gas is almost $60\%$ smaller. This 
indicates that a lower ram pressure is still efficient in stripping gas from the
disk, but part of the removed material is now bound to the satellite by the dark
halo gravity.  The simulations were computed using three different resolutions,
with $m_h/h_{disk}$ equal to 2, 1 and 0.5. The stripped mass
converges for $m_h/m_{disk} < 1$.

The dependence of ram pressure on the satellite orientation is not a
simple function of the angle between the disk plane and the orbital
velocity \citep{Vollmer} and can depend on the ram
pressure and the structure of the satellite \citep{Abadi, Quilis00, Schulz,
Marcolini}. In the simulations E1 and E2 the galaxy is moving edge on through
the hot gas. Comparing the amount of stripped mass of E1 and
E2 with the results of the corresponding face on runs H2 and L1, it is clear that
the stripping is not very sensitive to disk orientation. This implies that the
amount of gas in the Stream will only slightly depend on the initial inclination
of the LMC disk and on the precession eventually induced by tidal forces from
the MW.

\section{Interacting runs} \subsection{Orbital parameters} 
The morphology of the
Stream indicates that the orbit of the LMC is nearly polar and
counter--clockwise, as seen from the Sun, while kinematical data
\citep{vanderMarel02} imply that the present position at $\sim 50$ kpc from the
Galactic Center, is close to the perigalacticon.  Since the mass ratio between
the MW and its satellite is about 50:1, we expect that the effects of the
dynamical friction are not negligible on time scales comparable with the orbital
period. On the other hand, at each pericentric passage the LMC loses mass from
tidal stripping, increasing its decay time: dynamical friction becomes less
effective as the satellite approaches the Galaxy \citep{Colpi}.

\begin{figure} \epsfxsize=8truecm \epsfbox{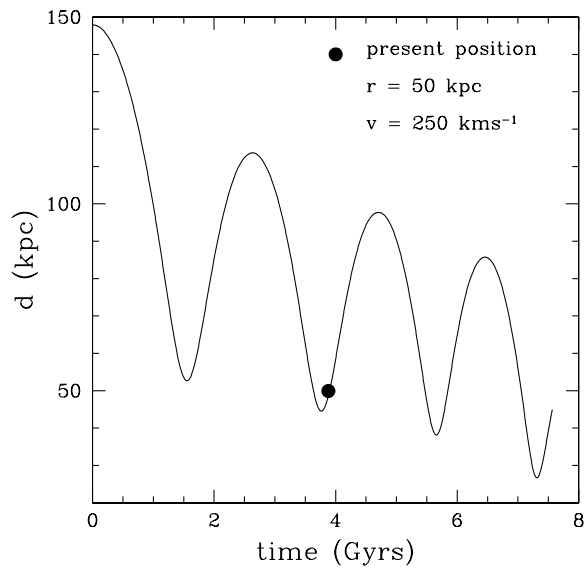} \caption{Orbital
separation for the MW -- LMC system} \end{figure}

Dynamical friction implies that we can not precisely make a backwards
integration of the LMC's orbit, therefore we make several low resolution test
simulations ($3\times 10^5$ particles) in order to find the correct 
starting point for the LMC 4 Gyrs in the past such that it ends up with
the present distance, velocity and orientation on the sky.
We find orbital parameters close to the ones obtained by
\citet{Gardiner94} and \citet{Gardiner96}. In particular the orbital plane is
perpendicular to the Galactic disk, the present eccentricity (defined as
$(r_{apo}-r_{per})/(r_{apo}+r_{per})$, with
$r_{apo}$ and $r_{per}$ apogalacticon and perigalacticon
distances) and the orbital period are respectively 0.44 and $\sim 2$ Gyrs. The
orbital separation between the MW and the LMC is plotted in Fig. 5 for 
the past 4 Gyrs and 4 Gyrs into the future.
The satellite is currently at 50 kpc from the Galactic center and is moving away
from the perigalacticon (at 45 kpc) with a velocity of $250\, \textrm{km}\,
\textrm{s}^{-1}$. From the plot it appears evident that the orbit is slowly
sinking, with each apogalactic distance $\sim 20 \%$ smaller than the previous
one. We will show in the next subsection that the change in the orbital
parameters due to dynamical friction strongly affects the tidal forces
and rate of ram pressure stripping.

In choosing the initial inclination and position of the line of nodes we make
the approximation that they do not change during the interaction. In particular
we adopt the values from \citet{vanderMarel02}: respectively $i = 34.7^{\circ}$
and $\Theta = 129.9^{\circ}$. This choice could partially influence the final
stellar structure of the disk, but not the ram pressure stripping process, as
seen in the previous section.

\subsection{The Stream}

The close interaction with the MW strongly perturbs the entire structure of the
satellite. In particular during the 4 Gyrs preceding the present epoch the LMC
loses $\sim80\%$ of its dark halo, and is currently 
tidally truncated at a radius of 11 kpc, consistent
with the values provided by analytic
calculations \citep{Weinberg00, vanderMarel02} and with observations of the
outer regions of the disk \citep{Irwin, vanderMarelII01}. 
The contribution of the dark matter halo to the final rotation curve within 8 kpc 
from the center decreases by $\sim 5 $ km s$^{-1}$ and the disk component 
predominates in the inner region (Fig. 6). 

\begin{figure} \epsfxsize=8truecm \epsfbox{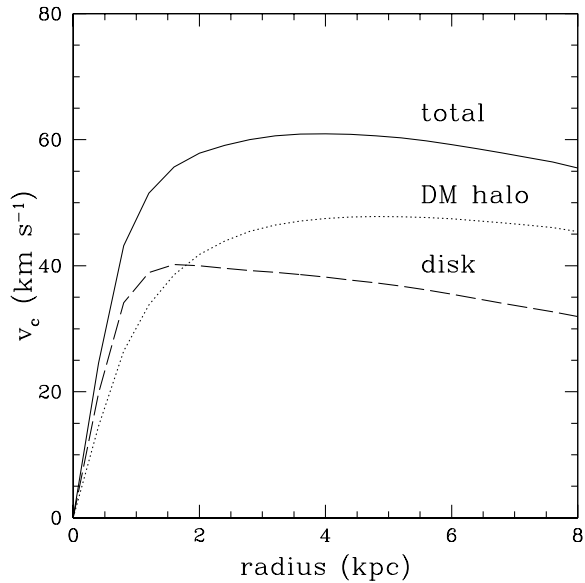} \caption{Final
LMC rotation curve.}  
\end{figure}

While all the stars within the tidal
radius are likely to be gravitationally bound, this is not true for gas. In
fact, part of the gas that lies within the tidal radius can have enough thermal
energy to escape from the potential well of the galaxy. Fig. 7 illustrates the
evolution of the baryonic mass as function of time. The dotted and solid curve
refers to the stellar and gaseus component respectively while the dashed curve 
represents the gas within the tidal radius which is actually bound to the LMC.  

\begin{figure} \epsfxsize=8truecm \epsfbox{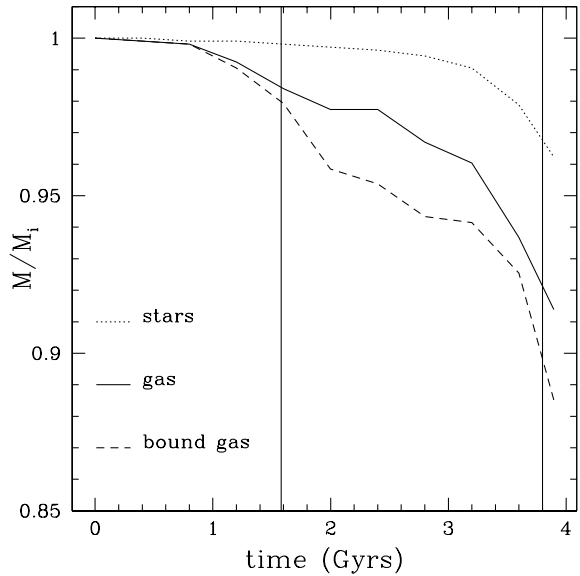} \caption{Fraction of
stars and gas bound to the satellite during the last 4 Gyrs. Both the gas
within the tidal radius and the bound gas are plotted. $M_i$ is the
initial mass of each disk component. The vertical solid lines indicate the
perigalactica. } \end{figure}

After two perigalactic
passages the satellite looses roughly $\sim 12\%$ of its gas and $\sim 4\%$ of
stars.  A pure tidal stripping model would remove similar
amounts of both stars and gas unless the initial gas disk was significantly more
extended. The larger amount of stripped gas is due primarily
to ram--pressure stripping: the rate at which gas is lost increases as the LMC
approaches perigalacticon (the solid vertical lines in the plot), corresponding
to higher densities of the diffuse halo gas and to higher velocities of the
satellite along the orbit.

Stars are stripped from the LMC only during the last perigalacticon, since the
previous perigalactic passage is too far from the MW to produce significative
tidal shocks and to perturb the stellar structure of the satellite. The
present stellar distribution (Fig. 8) indicates that a trailing and a small
leading arm are forming (with surface density  $\sim
3\times10^{-3} \textrm{M}_{\odot} \textrm{pc}^{-2}$), 
but most of the stars stripped from the disk are still
bound to the LMC and form a large spheroidal component.

\begin{figure} \epsfxsize=8truecm \epsfbox{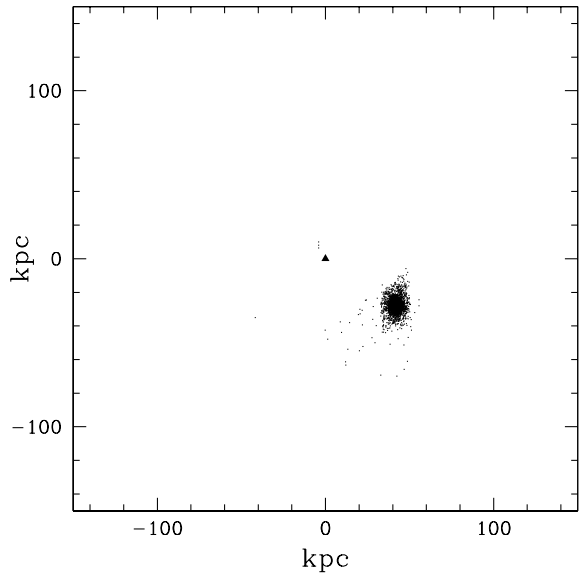} \caption{Present
day distribution of stars from the LMC disk in a plane perpendicular to 
the Galactic plane. } \end{figure}

\begin{figure} \epsfxsize=8truecm \epsfbox{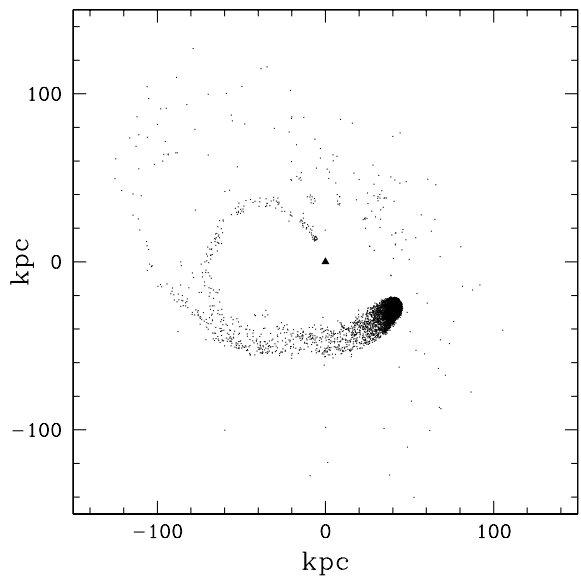} \caption{Final
distribution of gas from the LMC disk in a plane perpendicular to the Galactic
plane } \end{figure}

\begin{figure} \epsfxsize=8truecm \epsfbox{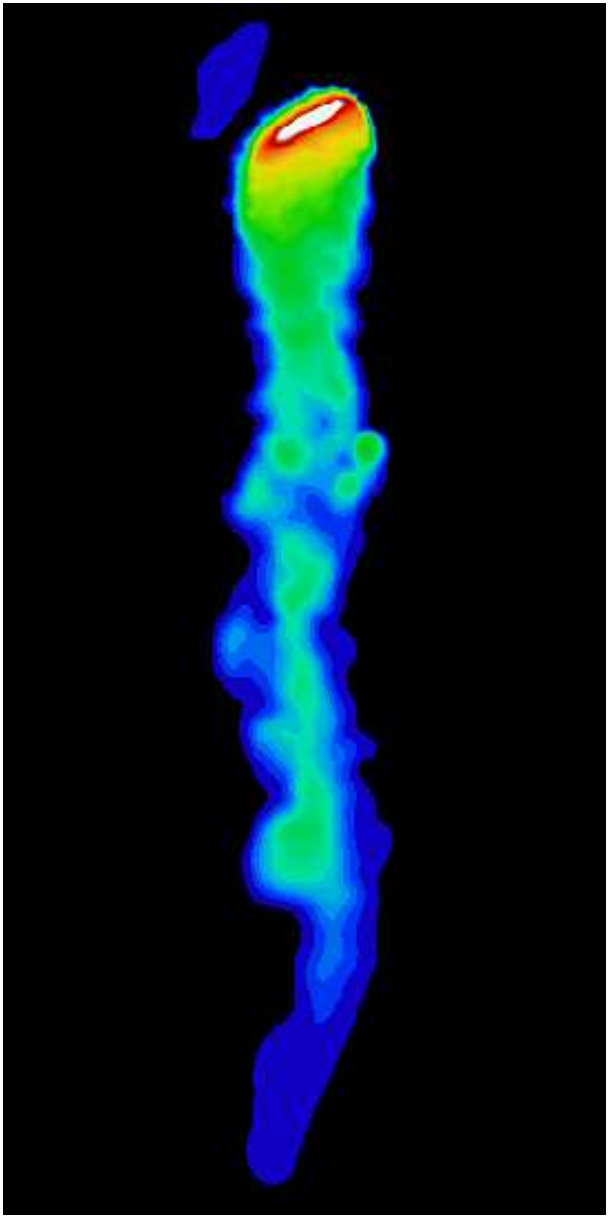} \vspace{0.5cm}
\caption{Hydrogen column density along the simulated stream. 
The intensity values are on
a logarithmic scale with white corresponding to the highest density regions
within the LMC disk and dark blue to $\textrm{n}_{\textrm{H}} \sim 10^{18}
\textrm{cm}^{-2}$.  } \end{figure}

\begin{figure} \epsfxsize=8truecm \epsfbox{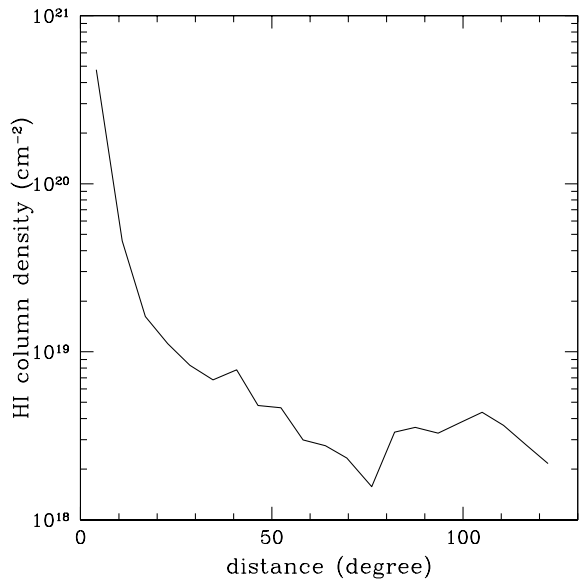} \caption{Hydrogen
column density along the Stream. The system of coordinates is chosen 
in order to have the LMC at 0 degrees. } \end{figure}

Contrary to the stars, gas is stripped during the first passage at 
perigalacticon, forming a continuous stream that lies in a thin plane (width
$\sim 25$ kpc) perpendicular to the disk of the MW (Fig. 9) and between
50 and 75 kpc from the Galactic center.
 Our simulations predict that the MS forms a great circle, but most of the gas
 is lost recently and it is now at an angular distance between
 $10^{\circ}$ and $120^{\circ}$ from the LMC (comparable with the range provided
 by observations). The mass of the stripped gas ($1.4 \times 10^{8} M_{\odot} $) is 
 of the same order of magnitude as the observed amount of neutral hydrogen
 in the Stream by \citet{Putman03} ($2 \times 10^{8} M_{\odot} $). 
 
 Fig. 10 represents the projection of the Stream as seen from an observer
 situated at the Galactic center who is looking toward the South Galactic
 Pole. The logarithmic colour scale indicates the column density
 gradient which gradually decreases along the length of the Stream.
 At a fixed distance from the LMC the structure of the Stream is not uniform but
 has clumpy and high density regions. 
 The decrease in density is by nearly two orders of magnitude along the Stream
 (Fig. 11), due to the combination of two factors. The gas close to the head of
 the Stream is pulled from the disk during the last perigalacticon, when the
 stripping rate is maximum, whereas the tip is $\sim 1$ Gyr old and corresponds
 to an apogalactic passage. Moreover, the tail is formed by the first material
 stripped from the outer and low density regions of the exponential disk. This
 decrease in column density is a remarkable success for the ram-pressure
 scenario since tidal models generically produce streams with surface densities
 that fall off much more slowly.

 The difference in the
characteristic time scales of hydrodynamical and gravitational forces explains
why a gaseous leading arm does not form through tidal stripping: gas is removed
quite early from the outer regions of the LMC disk that during the second
passage at the perigalacticon starts forming a leading arm feature. The
great circle of stripped gas actually appears on the sky as a leading
stream and could in principle reproduce observations of the leading arm
(Fig. 9).
This material was stripped from the satellite during the
first orbit and is falling to the Galactic center in the Northern Galactic
hemisphere, intersecting the actual position of the satellite. 

The total amount of unbound gas after 4 Gyrs is comparable with the gas 
stripped 
in one orbital time from an LMC
model in a test tube simulation with densities and velocities
characteristic of a perigalactic passage. This means that the tidal
forces acting on the satellite contribute to ram pressure processes:
as a consequence of the resulting asymmetrical potential and distorted
stellar disk, it is easier to strip more gas from the LMC's disk since
the gravitational restoring force is weaker. Thus even a low density
gaseus Galactic halo is able to remove a significant amount of gas
from the LMC.

\subsubsection{Adiabatic versus cooling runs}
We have presented
results from simulations that do not include radiative cooling. As discussed in
section 3, in this case the gas in the disk of the LMC will increase its
temperature as it is strongly compressed by the outer medium. In the
adiabatic regime the gas can only lose energy by expansion, which requires
a long timescale, and can become easily unbound due to this input of 
thermal energy. We have run a simulation with radiative cooling adopting
a standard cooling function for primordial gas (hydrogen and helium).
Since the gas quickly heats up to $\sim 10^5$ K, the peak of the
cooling curve, then cools down on a timescale much shorter than the 
dynamical time, shrinking in radius compared to its original location.
As a result a smaller gaseous disk results after a few cycles of heating and
rapid cooling which sits deeper in the potential well and as a result 
no gas is stripped. However, what we are witnessing is probably the
result of overcooling since we are neglecting 
the effect of star formation, namely heating from supernovae
feedback and radiation from massive stellar associations. The LMC in
particular shows widespread intense star formation across its disk
and bar. In reality a  multi-phase ISM will result that will be 
eventually close to the condition of pressure equilibrium with the outer 
medium and will not be able to cool so efficiently.
Until we can effectively model this multi-phase medium, we
believe that the results of the adiabatic simulations are closer to
reality.
\subsection{The LMC disk} The Galactic disk remains unperturbed during the
interaction, while the morphology of the satellite changes drastically. 
In particular we will focus on 
the changes in the stellar and gaseous structure of
the LMC. 

\begin{figure} \epsfxsize=7truecm \epsfbox{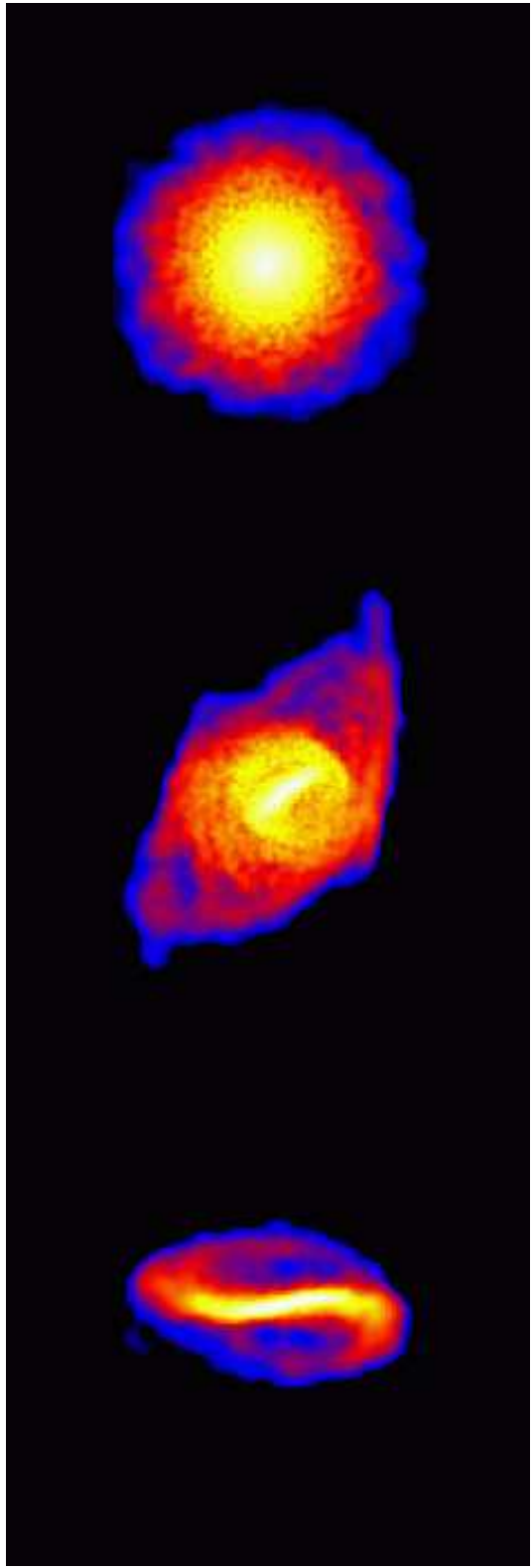} \caption{LMC stellar
surface density. Colours represent a logarithmic scale where white corresponds
to a density of $10^2 M_{\odot}\textrm{pc}^{-2}$ and blue to $10^{-2}
M_{\odot}\textrm{pc}^{-2}$. From the top to the bottom: initial conditions,
final face on and edge on projection. The bar and the strong warp are visible. } 
\end{figure}
 After 4 Gyrs from the beginning of the simulations almost the entire
stellar component is still bound to the satellite, but less than $90 \%$ of the
stars lies in the thin disk, which is surrounded by a stellar spheroid with a
radius of $\sim 11$ kpc (Fig. 12). The structure of this spheroidal component is quite
complex, since a strong warp that wraps $180 ^{\circ}$ around the LMC is
superimposed on the underlying low density stellar distribution. The warp forms
after the first passage at the perigalacticon and involves a large fraction of
the stellar particles. It extends up $\sim 5$ kpc out of the thin disk and is
aligned with the major axis of the satellite. The existence of a warp is
inferred by \citet{Kim} from the gas velocity field in the disk of the LMC, by
\citet{vanderMarelI01} and \citet{Olsen} using the apparent magnitude of
different classes of stars as a distance indicator. The diffuse spheroidal
component has an average surface density of $\sim 5 \times 10^{-2}
M_{\odot}\textrm{pc}^{-2}$, which becomes one order of magnitude larger if we
consider only the warp region. The warp deformation is partially directed along
the line of sight and could have implications for the microlensing results,
increasing the optical depth caused by self--lensing \citep{Zaritsky, Zhao}. The
LMC's disk appears elongated, with a mean ellipticity $\epsilon = 0.38$ (where
$\epsilon=1-q$ and $q$ is the axial ratio), 
 in a direction roughly aligned with the Galactic 
center and perpendicular to the Stream. These values agree very well with
  the observations provided by \citet{vanderMarelII01}. Within the inner 6--7 kpc 
 of the model LMC a
  strong and asymmetric bar forms after the first perigalacticon.

\begin{figure} \epsfxsize=20truecm \epsfbox{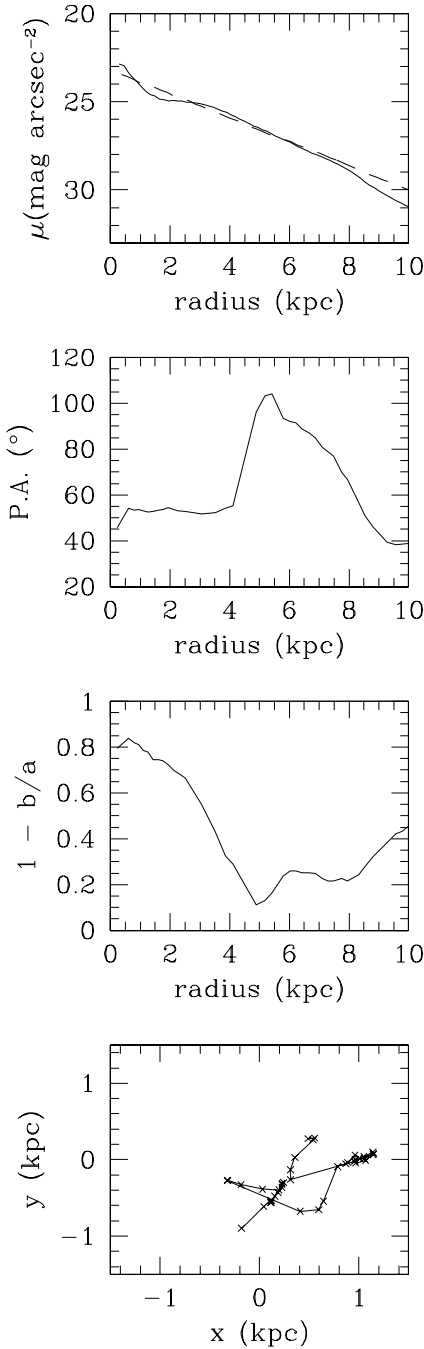} 
\caption {Photometric analysis of a face-on projection of the LMC disk. From the top to the bottom: B-band surface brightness, major axis position angle and ellipticity profiles. The last plot on the bottom represents the drift of the ellipse center in the plane of the disk.    
  } \end{figure}

The isodensity contours of the face-on disk were fitted with ellipses using the MIDAS task FIT/ELL3. The ellipse centers, ellipticity and major axis position angle were considered as free parameters of the fitting procedure. The results of this projected analysis are illustrated in Fig. 13. The first plot on the top (solid line) shows the B-band surface brightness profile of the present face-on LMC, for which the best fitting exponential curve has a scale length of 1.5 kpc. The dashed line is the original exponential disk, with a scale length of $\sim 1.7$ kpc. The second and the third plot illustrate the variation of the position angle and eccentricity as a function of radius. The position angle is almost constant in the central 4 kpc of the disk and presents a sharp twist beyond this radius, which corresponds to the transition between the bar region and the outer density contours characterized by ellipticities $\lsim 0.2$. At radii $\gsim$ 8 kpc the disk appears more elongated, with a further increase in ellipticities and a twist of $\sim 50^{\circ}$ in the position angle. The last plot on the bottom indicates the drift of the countour centers in the plane of the disk, which moves consistent with observations by almost 1 kpc.    

Fig. 14 shows the kinematic profiles of three different orientations of the LMC disk observed today: from the top to the bottom face-on, edge-on and end-on projections. The peak in the velocity dispersion in the central 6--8 kpc is mainly due to the bar formation, while the external regions appear to be dominated by an hot tidally perturbed component with $\sigma_p \sim 20$ km s$^{-1}$.

\begin{figure} \epsfxsize=8truecm \epsfbox{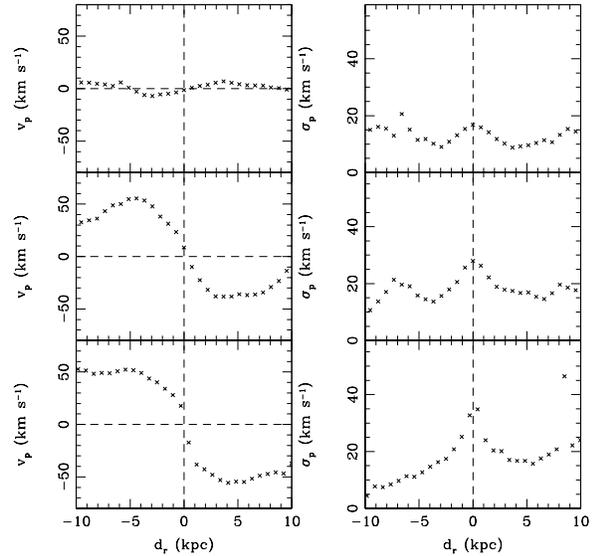} \caption{Projected velocity (left) and velocity dispersion (right) profiles for three different orientations of the LMC disk. From the top to the bottom: face-on, edge-on and end-on. } 
\end{figure}

\begin{figure} \epsfxsize=7truecm \epsfbox{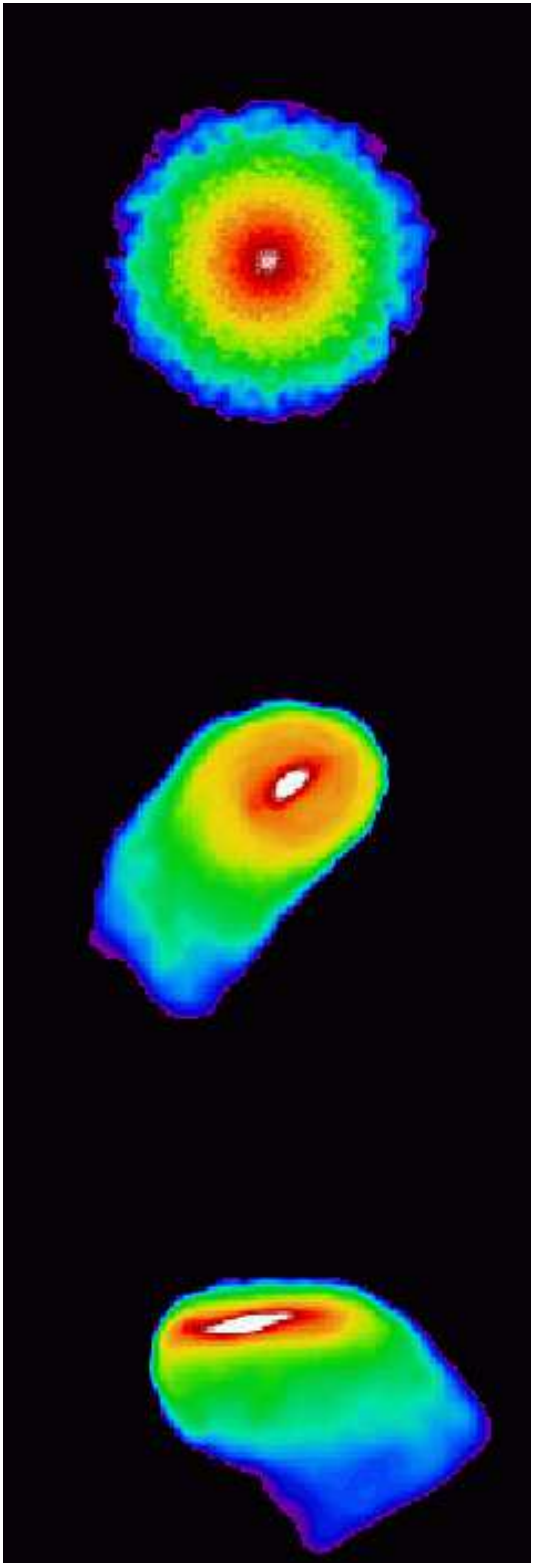} \caption{LMC gaseous
column density. Colours represent a logarithmic scale where white corresponds to
a density of $10^{22} \textrm{cm}^{-2}$ and violet to $10^{19}
\textrm{cm}^{-2}$. From the top to the bottom: initial conditions, final face on
and edge on projection. The bar and the first part of the Stream are visible. } 
\end{figure}

The gaseous disk is completely dominated by stripping processes (Fig 15). It retains a
symmetric structure only within the central 12 kpc, in connection with the thin
component of the stellar disk. Within this radius we do not observe a significative 
displacement between the center of the stellar distribution and the gaseus disk. 
The ram pressure stripping radius after 4 Gyrs is $\sim 6$ kpc, a factor of
three smaller than the initial disk radius and compatible with the results
provided by HIPASS data \citep{Putman98}.

\section{Conclusions}

We have carried out high resolution gravitational/hydrodynamical
simulations of the interaction between the LMC and Milky Way. As the
LMC spirals inwards towards the Galaxy it suffers ever increasing 
gravitational tidal forces and hydrodynamical stripping.
Our simulations cover the previous 4 Gyrs of the orbit of the LMC
such that at the final time it ends up at its correct location
within the Galactic halo and with the observed inclination.
We find that the combined effects of gravity and ram-pressure
stripping can account for the majority of the 
LMC's kinematical and morphological 
features and the morphology of Magellanic Stream. 

Our model of the hot gaseous halo places a total of $10^9M_\odot$ of 
ionised hydrogen within the dark matter halo of the Galaxy. The density
at 50 kpc is $\approx 8\times 10^{-5}$ atoms/cm$^3$. This is sufficient to remove
over $10^{8}M_\odot$ of gas from the LMC, close to the observed mass of the Stream.
The stripped gas forms a great circle, or a polar ring, around the
Galaxy, consistently with the recently discovered extension of the Stream into
the Northern Hemisphere by \citet{Braun}.
Less material was stripped during the pericentric passage 4 Gyrs
ago since dynamical friction has moved the Clouds closer today. 
Thus we can reproduce the observed decrease in column density along
the Stream. Very few stars are tidally removed from the LMC, but its disk
becomes severely warped due to the tidal interaction. This creates
a diffuse halo of stars within the LMC that may have been observed as 
self-lensing events. The disk of stars is elongated by 2:1 
oriented towards the Galactic centre in 
agreement with the observations of \citet{vanderMarelII01}.

Several improvements to our simulations could be made in order
to study this fascinating interacting system in more detail.
More detailed modelling of the star-formation and internal ISM of
the LMC is necessary in order to study the role of gas stripping
on star-formation, and the role of star-formation and the multiphase
ISM on the stripping process.
Including the interaction with the SMC which could supply a 
significant amount of gas to the Stream and may be 
responsible for the Magellanic bridge and the leading arm. 
Including a more realistic treatment of the gas physics in order
to study the detailed structure of the stripped gas and its
interaction with the halo gas as it infalls through the galactic halo.


\label{lastpage}

\end{document}